# Demystifying Progressive Web Application Permission Systems

Mengxiao Wang
Texas A&M University

Guofei Gu
Texas A&M University

## Abstract

Progressive Web Applications (PWAs) blend the advantages of web and native apps, offering features like offline access, push notifications, and installability. Beyond these, modern PWAs are increasingly granted system-level capabilities such as auto-start on login and shared context with native applications. However, their permission management remains poorly defined and inconsistently implemented across platforms and browsers.

To investigate these gaps, we developed Permissioner, a cross-platform analysis tool, and conducted a systematic study of PWA permissions. Our analysis uncovered critical issues of inconsistency, incompleteness, and unclear boundaries in permission enforcement, leading to various attacks including permission leakage, device identification, and Permission API abuse. We further examined why some browsers resist adopting more granular permission controls, identifying trade-offs involving usability, compatibility, and platform limitations. Through collaboration with browser vendors, several issues reported in our findings were acknowledged and resolved, notably by Firefox and Chrome. Our work highlights the urgent need for a unified, robust permission model for PWAs and provides actionable guidance toward achieving this goal.

## CCS Concepts

• **Do Not Use This Code → Generate the Correct Terms for Your Paper**; *Generate the Correct Terms for Your Paper*; Generate the Correct Terms for Your Paper; Generate the Correct Terms for Your Paper.

## Keywords

Do, Not, Us, This, Code, Put, the, Correct, Terms, for, Your, Paper

## 1 Introduction

Progressive Web Applications (PWAs) have emerged as a transformative technology, bridging the gap between traditional web applications and native mobile applications by leveraging core web standards such as service workers, web manifests, and HTTPS [54, 63].

PWAs enable richer and more persistent user experiences by offering offline access through service workers, real-time engagement via push notifications, and the ability to be installed on a user's home screen or desktop as standalone applications. Unlike TWAs, which are restricted to running in browser tabs, PWAs can operate in full-screen standalone windows with a native-like appearance, support background synchronization, and provide deeper OS-level integration. These capabilities have been significantly expanded through Google's Project Fugu [49], which introduced a series of new APIs aimed at closing the gap between web and native platforms. For example, after installation, a PWA can be automatically launched when the user logs into the operating system. PWAs can also be packaged as Android apps and published on various app stores, or share contexts such as text or files with other mobile apps. Details of these capabilities and the popularity of PWAs can be found in Section 2.

Beyond the aforementioned capabilities, PWAs have also gained access to a wide range of permissions that extend beyond those available to traditional web applications. For example, interfaces such as NFC, accelerometer, gyroscope, and the idle-detection API have been introduced to support mobile-like context awareness and continuous user interaction. These permissions enable PWAs to sense device motion, environmental context, and user inactivity, thereby unlocking sophisticated functionalities that were previously exclusive to native mobile apps. While these advancements offer clear benefits in terms of usability and user experience, they also introduce unique challenges regarding how permissions are defined, granted, and isolated across platforms and browsers. Critically, the behavior of these permissions remains dependent on browser-specific implementations. Such inconsistencies pose significant obstacles to achieving a uniform understanding and securing of PWA permissions across platforms. To address this issue, we developed a tool called *Permissioner*, which enables a systematic analysis of the PWA permission system, including the behaviors and inconsistencies of different Permission APIs across browsers.

Through our analysis of permission descriptors and policies, compared against those of mobile applications and traditional web applications, we uncovered critical inconsistencies across browsers and platforms. For instance, in the case of PWAs, permissions are managed on a per-origin basis rather than on a per-application basis. In contrast, mobile applications—such as those on Android—employ more fine-grained permission controls, including more specific options for geolocation access. Building upon these insights, we identified three novel types of attacks that demonstrate the security implications of the current PWA permission model. First, *PWA permission leakage attacks* exploit the absence of per-application permission isolation among PWAs that share the same origin. In such scenarios, a malicious PWA can silently inherit sensitive permissions that were previously granted to a benign PWA, without requesting additional user consent. This issue is further exacerbated by the fullscreen and app-store–like user interface of PWAs, which obscures origin information and enhances user trust. Additionally, a malicious PWA can misuse advanced permissions—such as those for NFC—to covertly read data from NFC tags or write malicious content to NFC cards. Second, *device identification attacks* leverage inconsistent default permission behaviors across different browsers and platforms. These inconsistencies give rise to side channels that facilitate robust fingerprinting of the user's device and browser, even in the absence of PWA installation or direct user interaction. Third, *Permission API abuse attacks* emerge from the underspecified and inconsistently implemented Permission API. In more severe cases, these design flaws enable malicious PWAs to crash the browser or trigger unintended behaviors, without requiring either installation or explicit actions from the user.

We further investigate why some browser vendors have not adopted per-app permission management for PWAs, despite its clear security benefits and partial implementation by others. Our analysis shows that this hesitation is not due to technical limitations, but rather trade-offs involving usability, backward compatibility, and perceived security risks associated with disrupting established permission models. To mitigate the identified attacks, we advocate for engaging browser vendors to resolve inconsistencies, standardize permission behavior, and enforce stricter isolation between PWAs under the same origin. Additionally, we propose dynamic reauthorization mechanisms for sensitive permissions—similar to iOS models—to align PWA permission management with their hybrid nature, improving security without compromising usability.

In summary, our contributions include:

- A comprehensive analysis of the PWA permission system, comparing it with mobile applications and traditional web applications. This analysis identifies key inconsistencies and security risks, and introduces *Permissioner*, a tool designed for systematic cross-platform analysis and testing of PWA permissions.
- The identification of three novel attack types: PWA permission leakage, device identification, and Permission API abuse. We propose corresponding mitigation strategies to strengthen permission isolation and management, supported by insights gained through collaboration with browser vendors.
- To support future research, all identified attacks, datasets, and tools presented in this paper have been open-sourced, providing the community with reproducible resources to replicate and extend our work [23].

## 2 Background and Motivation

This section introduces the foundational concepts of PWAs and compares them with other applications. It also discusses additional features unique to PWAs through deep integration. Additionally, we highlight the motivation for our research by identifying the unique characteristics of PWAs.

### 2.1 Progressive Web Applications and Their Capabilities

Before the emergence of Progressive Web Applications (PWAs), mobile users primarily accessed services via either Mobile Applications (MAs) or traditional web applications (TWAs). Each has distinct strengths and limitations. PWAs aim to combine the platform independence of TWAs with the rich functionality of MAs. We define these application types as follows:

- **Mobile Application (MA)**: A platform-specific app with deep device integration, offline access, and full hardware support. However, it requires separate development and maintenance for each operating system.
- **Web Application**: Runs in browsers and requires no installation. It includes:
  - **Traditional Web Application (TWA)**: Browser-dependent, cross-platform, and uses standard web APIs like the DOM and Geolocation. TWAs favor stable and widely supported APIs to ensure consistent performance.
  - **Progressive Web Application (PWA)**: Enhances TWAs with MA-like features such as installability, offline support (via service workers), and push notifications. A service worker alone qualifies a site as a PWA due to its background and offline capabilities.

PWAs leverage modern web APIs to offer rich functionality, although support varies across browsers (e.g., Sensor APIs are unsupported in Firefox and Safari). Compared to MAs, PWAs are inherently cross-platform, requiring only a single codebase, which reduces development cost and complexity [15]. They consume less storage by relying on browser environments rather than bundling full SDKs and resources. PWAs can also be accessed via URLs or distributed through app stores [25, 26].

To further differentiate PWAs from TWAs, a set of advanced capabilities becomes available only after installation [66]. These features enable deeper OS integration and native-like behavior. For example, *Run on Login* allows a PWA to auto-launch upon system login. While this improves usability, it also raises security concerns: a malicious PWA, once installed, can silently activate at login and operate without user awareness. The *File Handling API* permits PWAs to register as handlers for specific file types, integrating with the OS's *Open with* menu. Unlike traditional web apps—limited to file uploads via input fields—installed PWAs can open files directly from the system, granting them broader file system access and further blurring web-native boundaries. Other advanced features include *Web Share Target*, which enables PWAs to receive content from native apps via the system share menu. While these capabilities enhance the user experience, they also expand the attack surface. Without proper permission management, features like persistent background execution and system-level file access may be abused by malicious PWAs. Table 2 summarizes these install-time features.

As PWAs continue to gain access to system-level resources and privacy-sensitive interfaces, proper permission management becomes increasingly critical to prevent misuse. The expanded attack surface introduced by deeper OS integration underscores the need for robust, transparent, and user-controllable permission mechanisms [27, 32] Due to these advantages, many companies—including Instagram, Twitter Lite, Uber, and Starbucks—have adopted PWAs to meet growing mobile demands [24]. Notably, over 40% of Lyft users reportedly prefer the PWA over the full mobile app [24, 30, 36]. For more details about the popularity and adoption of PWAs, refer to Appendix A.1.

Table 1: Comparison of TWA, PWA, and MA Features (only distinct features are shown)

| Category | Detailed Features | TWA | PWA | MA |
|---|---|---|---|---|
| **Installation & Distribution** | Installability | ✗ | ✓ | ✓ |
| | Distribution | URL | URL, app store | app store |
| **User Experience Features** | Offline Support | ✗ | ✓ | ✓ |
| | Immersive Mode | ✗ | ✓ | ✓ |
| | Application Sharing | ✗ | ✓ | ✓ |
| **Device Access Features** | Push Notifications | ✗ | ✓ | ✓ |
| | Background Sync & Fetch | ✗ | ✓ | ✓ |
| | Hardware Access | ✗ | ✓ | ✓ |
| **Permission Features** | Permission Management Scope | browser-level | browser-level | system-level |
| | Permission Isolation Scope | origin-based | origin-based | application-based |
| | Permission Isolation Mechanism | browser sandbox | browser sandbox | application sandbox |

Furthermore, Table 1 compares key features across TWAs, PWAs, and MAs. While PWAs resemble MAs in terms of user experience and device access, their permission model remains more closely aligned with that of TWAs. This observation motivates a systematic examination of PWA permissions (see Section 2.2). A detailed feature-level comparison between PWAs and TWAs is provided in Appendix A.2. In this study, Android apps serve as the MA reference point for all comparative evaluations.

## 2.2 Motivation

Despite the increasing popularity of PWAs, we observe that, to date, there is no clear definition and understanding of what constitutes a PWA permission system. Unlike MAs or TWAs, where permission systems are well-established, PWAs lack a standardized permission management mechanism. As highlighted in Section 2.1, PWAs share certain functional similarities with mobile applications, particularly in terms of installation, user experience, and device access. However, their permission management approach aligns more closely with TWAs, suggesting that PWAs have inherited the TWA permission system. While this inheritance may simplify browser implementation, it introduces significant security and usability concerns that warrant further investigation.

Moreover, although PWAs are conceptually designed as an extension of web applications, they often exhibit platform dependence due to inconsistent browser implementations and handling mechanisms. This discrepancy challenges the promise of PWAs as a truly cross-platform solution. Understanding the unique aspects of the PWA permission system is therefore essential for ensuring security, usability, and platform independence in this emerging application model.

## 3 Methodology

Building on the background and motivation in Section 2, we formulated four research questions (RQs) to guide our investigation into the PWA permission system. These questions focus on understanding the system, identifying security risks, uncovering root causes, and exploring mitigation strategies:

- **RQ1 [Generic Understanding]**: What are permissions in PWAs, and how do they differ from or align with those in MAs and TWAs?
- **RQ2 [Risks from TWA-like Permissions]**: As PWAs evolve to adopt features similar to mobile applications, their permission model increasingly resembles that of TWAs. Will this TWA-like permission model introduce new security risks for PWAs?
- **RQ3 [Root Causes of Security Risks]**: To what extent are the identified security risks caused by browser-specific implementation decisions, technical constraints, or platform-level design philosophies?
- **RQ4 [Mitigation and Redesign]**: Given the identified root causes, how can the PWA permission system be redesigned to better mitigate security risks while maintaining usability?

To address these research questions, we conducted a comprehensive analysis of PWA permissions, focusing on their implementation and associated security implications. Beginning with an investigation into the current state of PWA permissions (**RQ1**), we gathered data from W3C documentation, browser source code, and developer forums to understand how permissions are managed. Empirical testing across major browsers, including Chromium, Safari, and Firefox, revealed variations in permission behavior. Using the Common Crawl dataset of PWAs [31], which includes 291,583 verified PWAs, we analyzed real-world implementations to uncover patterns in permission requests and usage. Our custom-built tool, *Permissioner*, enabled systematic categorization and assessment, exposing inconsistencies and alignments with other platforms.

Following this, we assessed the risks associated with the adoption of TWA-like permissions (**RQ2**). A comparative analysis of permission management mechanisms in PWAs, TWAs, and mobile applications revealed key differences in granularity, transparency, and isolation. Through threat modeling, we identified vulnerabilities such as privilege escalation and diminished user awareness, highlighting scenarios in which security risks may emerge as PWAs increasingly resemble native applications in behavior and capability.

To explore the underlying causes of these inconsistencies and risks (**RQ3**), we analyzed whether the observed permission behaviors result from differences in browser design, implementation strategies, or broader platform philosophies. Building upon our empirical findings, we examined the enforcement logic of the W3C Permission API across major browsers, with particular attention to Chrome's dynamic handling mechanisms—such as *permission prompt quieting* [51]—in contrast to the more static models in Safari and Firefox (e.g., permissions that persist until manually revoked or are requested on every visit). We then investigated whether aligning PWA permissions with mobile app policies (e.g., session-based or revocable models) is hindered by technical limitations, constrained by usability trade-offs, or influenced by platform-level priorities. This analysis allowed us to identify the root causes behind permission inconsistencies and evaluate the feasibility of adopting mobile-style permission governance in web environments.

Table 2: Selected System-level Features Only Available to Installed PWAs

| Feature | Description |
| --- | --- |
| Run on Login | Automatically launch the PWA when the user logs into the OS. |
| getInstalledRelatedApps | Detect whether related native apps (e.g., Android or Windows) are already installed. |
| Trusted Web Activity | Package the PWA as an Android app for distribution on Google Play. |
| File Handling API | Register the PWA to open specific file types via OS-level *Open with* menus. |
| Web Share Target | Receive shared content (text, files) from other apps through OS sharing. |

Building on these insights, we proposed and evaluated mitigation strategies aimed at improving the security posture of PWA permissions (**RQ4**). Specifically, we explored the feasibility of redesigning the permission system to incorporate session-based lifecycles, finer-grained permission scopes, and improved user-facing transparency. These proposals were developed in line with modern usability principles and web security models. We implemented prototype solutions and conducted controlled user studies to evaluate their effectiveness in enhancing user awareness and minimizing over-permissioning. Through collaboration with browser vendors, we assessed the practicality and deployability of these strategies, demonstrating their potential to strengthen the PWA ecosystem without compromising usability.

## 4 Understanding PWA Permissions

To address RQ1, we began by reviewing official documents, community forums, and browser developer guides to understand the definition of PWA permissions. Our key finding indicates that no explicit definition of a PWA-specific permission system is provided by any browser vendor or the W3C. The W3C documentation mentions the *Permission API* [70] for web applications, leading us to hypothesize that browser vendors employ the same permission management system for PWAs as for TWAs.

We curated a verified dataset of PWAs based on the Mozilla definition [56], building on prior work [31], and ensuring that all included entries met the installation criteria specified in the PWA standards. Each PWA in our dataset contains a manifest file with valid fields—such as *display*—that are essential for installation and functionality. This step addresses inaccuracies in previous studies, which often included entries that did not fully comply with PWA standards. Details of the dataset and experimental setup are provided in Appendix A.3.

Table 3: Top 10 Permission APIs Used by PWAs (By Count)

| Permission API | Invocable | Prompted | Count |
| --- | --- | --- | --- |
| Clipboard Write | ✓ | ✗ | 32,135 |
| Clipboard Read | ✓ | ✓ | 24,753 |
| Geolocation | ✓ | ✓ | 11,350 |
| Background Sync | ✓ | ✗ | 10,456 |
| Notifications | ✓ | ✓ | 8,691 |
| Fullscreen | ✓ | ✗ | 5,336 |
| Microphone | ✓ | ✓ | 2,970 |
| Camera | ✓ | ✓ | 2,959 |
| Storage Access | ✓ | ✗ | 673 |
| Display Capture | ✓ | ✗ | 539 |

Through this methodology, we identified a total of 291,583 installable PWAs. To conduct a comprehensive security risk analysis, we focused on collecting and analyzing code related to permission usage. Specifically, we targeted the use of the Permissions API [21] and associated Web APIs such as navigator.geolocation. Based on the official specification [21], we manually enumerated common usage patterns of these permission-related Web APIs. This included both direct invocations (e.g., navigator.geolocation) and indirect patterns observed in third-party JavaScript files embedded via <script src=" ">. To ensure high coverage, we compiled a set of regular expressions that match various coding styles and invocation structures of these APIs. Using Puppeteer [61], a headless browser automation tool, we developed a crawler to simulate real user interactions and fully load JavaScript execution contexts. Unlike static analysis, our crawler inspects both inline scripts embedded in the main HTML as well as third-party scripts dynamically loaded during runtime. This approach enables us to detect permission requests that are issued via external scripts, which are common in advertising, analytics, or utility libraries. In addition to identifying permission usage, we also examined the structure of each origin to determine whether it hosts multiple PWAs. For each detected PWA, we scanned for other URLs within the same origin that are installable and contain a valid manifest.json. If multiple start_url values under the same origin point to different application entry points, we consider the origin to contain multiple distinct PWAs, commonly referred to as standalone PWAs, under a shared origin. To avoid overloading target servers, we imposed a 30-second timeout per website during crawling. This ensures responsible crawling behavior and avoids inadvertently degrading website performance. While this conservative timeout may lead to missing some slower-loading or deeply nested PWAs, it strikes a balance between comprehensiveness and ethical data collection.

Table 3 presents the ten most commonly used Permission APIs among PWAs (a complete list is available in Table 15), highlighting whether each API is invocable and whether it prompts users for explicit consent. Here, *Invocable* indicates whether the API can be programmatically executed, while *Prompted* refers to whether user interaction is required to grant access. Notably, the most frequently used permission is clipboard-write, appearing in 32,135 PWAs. However, as this permission does not trigger a user prompt, it allows data to be written to the clipboard without explicit user consent. The top five prompted permissions are clipboard-read, geolocation, notifications, microphone, and camera. For further analysis related to prompted permissions, we focus on these five examples, as they are widely utilized compared to other permissions. A detailed comparison of all permission descriptors will be discussed in Appendix A.4.

To further explore the scope of permissions, we analyzed the source code of major browsers, including Chromium [19], Firefox [22], and WebKit [20]. Our review revealed that browser vendors define permission descriptors, which play a critical role in the implementation of the Permission API. A permission descriptor is a structured definition of a browser permission, encompassing properties such as its name (e.g., *geolocation*). We found that Chromium-based browsers support 32 permissions [19], encompassing the 15 permissions in iOS-based browsers [20] and the 9 in Firefox [22]. Chromium-based browsers provide the most extensive set of descriptors, while the others rely on subsets. Our study focuses on Chromium-based browsers and Android devices. Although we conducted experiments on iOS, all iOS browsers share the same engine and exhibit similar behavior. Furthermore, iOS support for PWAs is akin to bookmarks rather than standalone apps. Thus, subsequent sections primarily focus on Android browsers. A complete list of permission descriptors is provided in Appendix A.6. Source code analysis revealed that permission descriptors correspond directly to the Permission API, which the official documentation [70] corroborates by clarifying the relationship between specific descriptors and their implementations. While permission descriptors define permissions at a low level, the Permission API provides a developer-friendly interface for querying or requesting permissions, relying on these descriptors to manage states and process queries.

## 4.1 Permissioner

Through preliminary experiments, we identified the relationship between permission descriptors and the Permission API. However, our observations revealed that PWA permissions evolve over time. For instance, in the Chromium source code, the *window-placement* permission was merged into *window-management*, while the *accessibility-events* permission was entirely removed [19]. Furthermore, some permissions, such as *web-app-installation*, were unsupported before the commencement of our experiments [19]. This dynamic nature highlights the importance of validating permissions' usability and understanding their real-world adoption by PWAs. Notably, the official documentation [70] regarding Permission API support is often outdated and subject to frequent changes, as demonstrated in Appendix A.6.

To bridge the gap between outdated documentation and evolving permission management implementations, we developed a tool named **Permissioner**. This tool identifies permissions accessible to PWAs and determines which permissions are actively utilized in practice. Figure 1 illustrates the workflow of **Permissioner**, which comprises two main components: *Permission API Identification* and *Dynamic Permission Analysis*.

The *Permission API Identification* component catalogs the APIs managing permissions by examining browser source code and documentation. It maps permission descriptors to their corresponding APIs, establishing a foundation for identifying compatibility and behavioral inconsistencies across browsers.

The *Dynamic Permission Analysis* component evaluates permissions through simulated user interactions to determine their practical applicability in real-world scenarios. This process verifies the functionality of permissions, including their availability and implementation in PWAs. Once the invocability of permission APIs is confirmed, the analysis extends to examining permission isolation policies and granting mechanisms. This focus is essential for identifying potential security vulnerabilities and inconsistencies in cross-browser behavior.

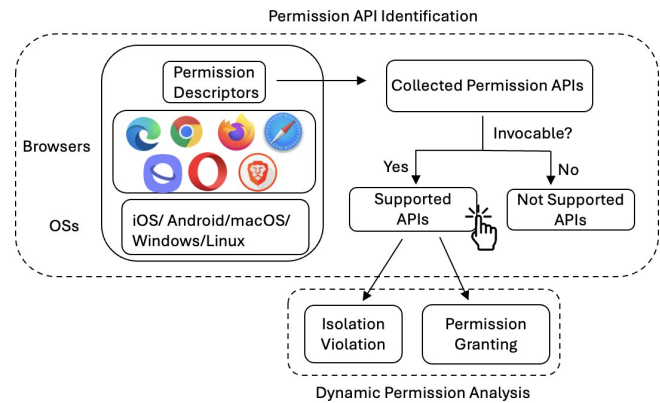

**Figure 1: Workflow of Permissioner**

## 4.2 Permission Descriptors Comparison between TWA, PWA, and MA

Browsers do not differentiate between the permission systems of PWAs and TWAs, leading to ambiguity and overlap between the two models. Permission descriptors, which serve as identifiers for supported permissions in the Permission API, are shared between PWAs and TWAs without explicit differentiation. This unified model overlooks the unique capabilities and risks associated with PWAs. This lack of separation introduces security concerns. For instance, PWAs can operate in the background using service workers, making certain permissions, such as microphone or camera access, more powerful than in TWAs. Permissions related to background operations (e.g., *background-fetch*, *background-sync*) are entirely dependent on service workers, which are unique to PWAs. These differences highlight the enhanced functionality of PWAs but also expose potential vulnerabilities, as powerful permissions in the context of PWAs are not subjected to stricter control or isolation.

To address these issues, we categorized permissions based on their role in enhancing PWA functionality to align more closely with native mobile applications. Table 4 presents this categorization. Permissions highlighted in green provide a mobile-like experience in PWAs, granting access to key device hardware and system features that are integral to mobile applications. These permissions enable richer interactions, such as sensor-based motion tracking, NFC-based interactions, and background notifications, which significantly improve user experience compared to traditional web applications.

**TWA and PWA Permission Descriptor Categorization.** As shown in Table 4, while multiple permissions exist across platforms, only camera, geolocation, and microphone are universally supported across all tested browsers and platforms in our experiments. Compared to TWAs, PWAs exhibit lower cross-platform consistency due to browser-specific implementations of the Permission API. Different browsers define and enforce permission handling with varying levels of granularity and default behaviors. Our categorization is conceptually motivated to better distinguish mobile-enhancing permissions rather than solely reflecting how different browsers classify them. This approach allows a clearer differentiation between permissions that improve the PWA mobile experience and those that remain more generic or desktop-oriented. While browsers may differ in their permission structures, our classification helps contextualize security risks and usability concerns, offering a more structured understanding of PWA permission management.

In contrast, related work [58] includes notifications as a TWA-specific permission. However, this overlooks Opera, where notifications are rejected by default, unlike other browsers where they are prompted. Furthermore, notifications manage both the Notification API and the Push API, with the latter being exclusive to PWAs. The extensive additional permissions in PWAs underscore that their permission system is more complex and often underestimated compared to TWAs.

PWA permission systems differ from those of MAs, lacking standardized definitions and granular controls. Improving transparency by implementing permission descriptors in PWA manifests and offering more explicit prompts could enhance user control and reduce overprivileged access. Since PWAs provide a mobile-like experience and Android treats PWAs installed from Chrome as real applications [37], we compared the PWA permission system with the Android permission system. According to Android documentation, 308 permissions are defined in *Manifest.permission* [13]. The Android system enforces developers to declare permissions in their manifest, enabling security researchers to build permission maps to identify overprivileged applications [13]. The Android permission system categorizes permissions into three types: runtime, install-time, and special. For runtime permissions, Android prompts users to grant or deny requests explicitly through a dialog box. In contrast, PWA permissions lack such uniform definitions due to browser-specific permission descriptors. Chromium-based browsers like Brave, Edge, and Opera support fewer API permissions compared to Chrome, leading to inconsistencies across platforms. For install-time permissions in Android, users are shown a list of required permissions for verification before installation. However, in PWAs, allow-by-default permissions resemble install-time permissions but are not explicitly presented to users. We propose mimicking the Android permission system for PWAs by declaring permission descriptors in PWA manifests. Some PWAs, such as Google Maps, already implement this feature by declaring *gcm* permissions in their manifests [11]. Enforcing this approach across browsers could make the PWA permission system more transparent and reduce overprivileged access.

Another notable difference between the PWA and Android permission systems is the granularity of permissions. For instance, Android distinguishes between *ACCESS_COARSE_LOCATION* and *ACCESS_FINE_LOCATION*, granting approximate and precise location permissions, respectively. Implementing a similar mechanism in PWAs could better protect user privacy. Additionally, PWA permission prompts offer between two and four options (as detailed in Table 16), whereas Android permission prompts typically provide three options. Related works [40] indicate that when only *allow* and *deny* options are presented, 84% of participants grant permissions, often out of fear they cannot enable them later. Introducing additional options, such as *allow this time*, could improve user control and reduce unnecessary permissions, making the PWA permission system more user-friendly.

## 4.3 Permission Policy Comparison Between PWAs and MAs

This section compares *permission isolation* and *permission granting mechanisms* between PWAs across different browsers and MAs on Android. Since PWAs directly adopt the permission policies of TWAs, we exclude a comparison between PWAs and TWAs, as their mechanisms are fundamentally identical. Instead, our analysis focuses on the distinctions between PWAs and MAs on Android, highlighting the challenges PWAs face due to their reliance on browser-defined policies. PWAs present notable issues with permission sharing and management. Permissions may be shared across different instances or browsers, introducing complexities and increasing security risks. In contrast, Android applications enforce stricter isolation and finer-grained permission granting mechanisms, offering users greater transparency and control.

### A. Permission Isolation Comparison

**Permission Isolation in Android.** Permission isolation in Android is a security feature that isolates applications from each other and the underlying system. This mechanism prevents applications from accessing sensitive resources or performing restricted operations without explicit user permission. Android achieves this by sandboxing applications, ensuring each operates independently without interfering with others.

**Permission Isolation in PWAs.** The permission isolation mechanism in PWAs differs significantly from that of native applications. To analyze this, we define distinct PWAs based on their unique application identifiers (App IDs) [70]. PWAs installed from the same URL but through different browsers should be treated as separate entities, as they run within different browser environments and should not share permissions. PWAs originating from the same

Table 4: Categorization of TWA and PWA Permission Descriptors (Green: Permissions Enhancing Mobile-like Experience)

| Category | Permissions |
| --- | --- |
| Sensor | accelerometer, ambient-light-sensor, gyroscope, magnetometer |
| Hardware Access | nfc, camera, microphone, speaker-selection |
| Clipboard & Data Access | clipboard-read, clipboard-write, persistent-storage, storage-access, top-level-storage-access |
| Notifications & Background Processing | background-fetch, background-sync, periodic-background-sync, notifications, push |
| Location & Environmental Awareness | geolocation, idle-detection, screen-wake-lock, system-wake-lock |
| Window & UI Management | fullscreen, pointer-lock, window-management |
| Payment & Authentication | payment-handler, midi |
| Desktop-related | display-capture, local-fonts, captured-surface-control, keyboard-lock, web-app-installation |

domain but with different App IDs can share permissions without explicit user consent. This behavior allows less secure PWAs to inherit permissions from more trusted ones, posing significant security risks, as discussed in Section 5.1. In contrast, Android applications isolate permissions per application, requiring independent user approval for each. This strict isolation enhances security and prevents permission inheritance.

**Permission Sharing in PWAs.** For PWAs hosted on different browsers, they are considered distinct applications, as Android browsers do not share permissions for the same PWAs. However, on iOS devices, browsers such as Safari, Firefox, and Chrome share geolocation permissions for the same PWAs, as reported in related works [58]. These interconnected permissions increase the attack surface, as permissions granted in one browser may cascade to others. Another inconsistency lies in the isolation of permissions between PWA sessions and browser sessions. On Android and desktop platforms, these two sessions utilize the same permission system. In contrast, on iOS devices, except for geolocation permissions, sessions are isolated. This behavior creates security gaps, as permissions can persist across contexts even after clearing browser settings. For instance, on Chrome and Opera, notification permissions persist even when browser settings are cleared, provided the PWA remains installed.

### B. Permission Granting Mechanism Comparison

This section compares the permission granting mechanisms of PWAs and MAs (Android), focusing on user interactions with permission prompts and the implications of default permissions. MAs (Android) generally employ a more fine-grained approach compared to PWAs, offering better user control and transparency.

**User Interaction with Permission Prompts.** Permission prompts in PWAs vary across browsers, resulting in inconsistencies in user experience. While MAs (Android) provide clear options such as *allow only this time*, *allow while using the app*, or *deny*, PWAs often lack similar granularity. For instance, some Android browsers, like Brave, offer extended options (e.g., *allow for 24 hours* or *allow permanently*), but such flexibility is inconsistent across platforms. This disparity can confuse users and increase the likelihood of unintentionally granting permissions. In contrast, Android native applications enforce stricter boundaries by explicitly associating permissions with app sessions. This approach ensures users understand the scope of granted permissions, reducing the risks of inadvertent or persistent permissions.

Table 5: PWA Permission Default Status in Various Android Browsers.

| PWA Permission | C | S | F | E | O | B |
| --- | --- | --- | --- | --- | --- | --- |
| accelerometer | g | g | - | g | g | d |
| background-fetch | g | g | - | g | d | g |
| background-sync | g | d | - | g | g | d |
| gyroscope | g | g | - | g | g | g |
| magnetometer | g | g | - | g | g | g |
| periodic-background-sync | g | d | - | d | p | d |
| screen-wake-lock | g | g | - | g | d | g |
| storage-access | g | p | p | g | - | - |

Browsers: C (Chrome), S (Samsung Internet), F (Firefox), E (Edge), O (Opera), B (Brave). p: Permission requested with prompts; d: Permission denied; - : Not supported; g: Permission granted.

**Default Permissions and Risks.** Certain permissions in PWAs are granted by default, raising significant security and privacy concerns. A comprehensive list of PWA permission default values is provided in Table 10, while Table 5 highlights permissions with inconsistent default values across different browsers on Android devices. Permissions such as background-fetch, background-sync, and periodic-background-sync are often granted by default, enabling background tasks without explicit user approval. While convenient, these permissions pose potential risks, including persistent tracking and unauthorized background activity. Sensor APIs like accelerometer and gyroscope further exacerbate these risks. Although similar permissions are also granted by default in Android native apps, they are better studied and mitigated in the native app context. For instance, sensor APIs can detect user movement patterns or device usage, posing significant privacy threats [43, 69].

However, these issues remain underexplored in the PWA context, leaving room for attackers to exploit these sensors for unauthorized data collection. For a detailed analysis of permissions, refer to Appendix A.5.

## 5 Assessing Risks from TWA-like Permissions

Based on the intriguing findings from RQ1, this section introduces three attacks targeting the PWA permission system: the PWA Permission Leakage Attack, the Device Identification Attack, and the PWA API Abuse Attack. For each attack, we developed a comprehensive threat model and conducted a practical analysis by examining real-world PWAs. Our findings reveal that similar vulnerabilities exist in TWAs and other systems; however, attacks on PWAs are significantly more stealthy, powerful, and persistent compared to those in other systems.

### 5.1 PWA Permission Leakage Attack

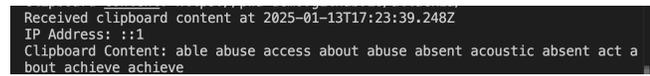

**Figure 2: Screenshot of PWA Permission Leakage Attack exposing crypto seed and IP.**

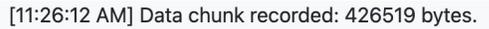

**Figure 3: Screenshot of PWA Permission Leakage Attack with unauthorized video recording.**

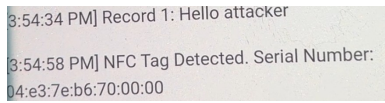

**Figure 4: Screenshot of PWA Permission Leakage Attack utilizing NFC.**

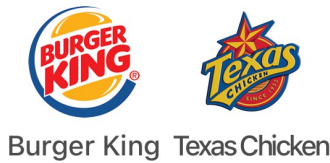

**Figure 5: Screenshots of different PWAs using different icons but sharing the same permissions**

In recent years, a growing trend has emerged in which multiple PWAs are hosted under the same origin by independent developers. Platforms such as *chatfood.io* [5] and *crazygames.com* [7] exemplify this model, allowing developers to upload their code to shared remote servers without the need to maintain dedicated infrastructure or build comprehensive websites. This approach has gained popularity due to its cost-efficiency, enabling developers to reduce operational overhead by eliminating the complexities of server management and website development.

However, this convenience introduces notable security risks. A critical concern is the unintended sharing of permissions granted under the same origin across different PWAs. As illustrated in Figure 8, two distinct PWA games hosted on the same origin share identical permissions. For example, if a user interacts with a benign PWA and grants it permissions, such as access to the clipboard, a malicious PWA hosted under the same origin could exploit these permissions without requiring additional user consent. As shown in Figure 5, both Burger King and Texas Chicken are hosted under the same origin, order.chatfood.io, and share permissions such as geolocation. However, from the user's perspective, these PWAs can only be distinguished by their icons on the home screen or when opened, since the URL is not visible in the PWA interface. This vulnerability underscores the need for enhanced permission management mechanisms to mitigate such risks.

This problem is not observed in mobile applications because mobile platforms enforce strict isolation policies for apps distributed via app stores. Each mobile app operates within its own isolated environment, ensuring that permissions granted to one app cannot be accessed by another. In contrast, the shared origin model of PWAs creates a scenario where such isolation is absent, leading to PWA permission leakage.

In this attack scenario, we assume that the attacker does not possess complete control over all PWAs within the shared origin and does not control the origin itself. Instead, the attacker has the capability to upload their own PWA to the shared origin. For example, on platforms like *crazygames.com*, developers are allowed to upload HTML pages as games, all hosted under the same origin. This setup provides an entry point for malicious actors to exploit shared permissions.

**Table 6: Comparison of PWA Permission Leakage Attacks and TWA-Based Attacks**

| Category | PWA Permission Leakage Attacks | TWA-Based Attacks [71] |
|---|---|---|
| Stealth | Hidden origins<br>Hidden in app store | Visible origins<br>Through URL access |
| Capabilities | Broader permissions (e.g., NFC, clipboard)<br>Background context monitoring | Standard web permissions<br>No background context awareness |
| Persistence | Offline functionality<br>Higher engagement rate | No offline functionality<br>Low engagement rate |

As demonstrated in Figure 9, we define a benign PWA, referred to as *PWA1*, which is a game application capable of requesting various permissions, including microphone, camera, notifications, clipboard, NFC, and geolocation. These permission requests are assumed to be reasonable within the context of *PWA1*, and it is expected that users would grant them. In contrast, the attacker-controlled PWA, referred to as *PWA2*, is designed with its own service worker that continuously queries whether these permissions have been allowed. Once the permissions are granted to *PWA1*, the service worker in *PWA2* detects this and immediately sends a push notification to entice the user to open *PWA2*. Upon opening, *PWA2* exploits the shared permissions to directly access sensitive user data, such as geolocation, NFC context (e.g., Figure 4), clipboard content (e.g., crypto seed, as shown in Figure 2), and recorded video, which can be stored in *IndexedDB*. Even if *PWA2* is later closed, the service worker remains active in the background, enabling it to process video data and transmit it to a remote server (e.g., Figure 3). Since *PWA2* is controlled by the attacker, it can customize push notifications to lure the user into opening *PWA2*. For instance, a notification leveraging *crypto news* can entice the user. This attack can also incorporate context-awareness; the service worker in *PWA2* monitors the user's IP address to estimate their approximate location and status, adapting the timing and content of push notifications based on changes in IP and time to deliver highly personalized and effective attacks [23]. Once the attacker acquires user profiles, these can be monetized through various illicit means, such as selling the data for illegal activities, conducting phishing attacks, or stealing cryptocurrency using the extracted crypto seed. Importantly, this entire process operates seamlessly without the user's awareness, relying solely on the user clicking the enticing push notification to unknowingly execute the malicious intent.

Our empirical analysis identified 12,487 origins hosting multiple PWAs, with each origin containing at least two PWAs. Notably, one origin hosts as many as 8,000 distinct PWAs. As shown in Table 7, 378 origins share geolocation permissions, 324 share notification permissions, 12 share clipboard read permissions, 85 share media permissions (microphone and camera), and 1 origin shares NFC permissions. A comprehensive security experiment was conducted on a representative case, *crazygames.com*, following explicit consent from the respective PWA developers, who acknowledged and validated the identified permission-sharing vulnerabilities. The experiments were strictly confined to scenarios observable in developer mode, and both attacker and victim interactions were systematically simulated using a benign PWA.

It is critical to emphasize that these vulnerabilities are not attributable to individual PWA developers but rather to the inadequate enforcement of permission isolation and separation policies by browser vendors. Each PWA is associated with a unique application identifier and is frequently managed by different developers, rendering assumptions of inherent trust among them inappropriate. While not all of the 12,487 origins currently exhibit permission-sharing behavior, those that may request permissions in the future remain vulnerable to similar exploitation. These findings underscore the pressing need for browser vendors to reevaluate and enhance the implementation of permission isolation mechanisms to mitigate these systemic risks effectively.

TWAs also encounter similar issues where permissions are shared under the same origin, even when developed by different developers. Takuya et al. [71] proposed a concept called web rehoisting. Websites like *proxysite.com* [8] host different websites under the same origin. As illustrated in Figure 10, even the same websites during the first and second visits may not guarantee the same origin for the rehost. However, such websites explicitly display their URLs, enabling users to identify the origin they are interacting with.

Table 7: Number of origins sharing permissions across multiple PWAs.

| Permission Type | # of Origins Sharing Permissions |
|---|---|
| Geolocation | 378 |
| Notification | 324 |
| Clipboard Read | 12 |
| Media | 85 |
| NFC | 1 |

As described in Table 6, PWA permission leakage attacks exhibit greater stealth, more advanced capabilities, and increased persistence compared to TWA-based attacks. From a stealth perspective, PWAs can be installed directly from app stores, such as the Microsoft App Store, which installs PWAs locally without encapsulating their permissions. Our experiments reveal that this process prevents users from identifying the URL, making the attack more covert. Additionally, PWAs operating in fullscreen mode inherently lack a visible URL, further enhancing their concealment. Regarding capabilities, PWAs have access to a broader range of permissions, such as NFC (detailed with real examples in Appendix A.8.2) and clipboard access, enabling a wider scope of operations. Furthermore, PWAs utilize service workers, allowing them to execute code in the background or even after being closed. For instance, service workers can detect changes in user states, such as permission changes or IP address updates, and leverage this information to send more targeted push notifications, encouraging user re-engagement. In terms of persistence, PWAs can store exfiltrated data in IndexedDB while offline and transmit it to a remote server once the device regains connectivity. As demonstrated in Appendix A.8.3, the click-through rate (CTR) for PWAs is significantly higher than that of TWAs, attributed to features like push notifications and enhanced functionality. Consequently, PWAs exhibit substantially higher re-engagement rates compared to TWAs.

## 5.2 Device Identification Enhancement

Device identification is a technique used to determine the specific hardware platform a user is operating on, such as a desktop computer, an iOS device, or an Android device, as well as the specific browser being used. This capability allows attackers or legitimate entities to gain insights into user behavior, track devices across sessions, and target users, or even conduct phishing attacks using browser and platform-specific knowledge. Our approach acts as a *valuable supplement* to existing device/browser fingerprinting techniques by introducing a more robust feature set that remains effective even when browser fingerprints fail or become unreliable. This technique serves multiple purposes, including application optimization, targeted advertising, enhanced security, and system protection. By utilizing browser fingerprints, applications can dynamically adapt content for Android or iOS users, thereby improving the user experience even without specific device model information. Additionally, advertisers leverage browser information to deliver personalized advertisements, such as specific promotions tailored for Samsung Internet users. Security systems incorporate device identification information into authentication processes, detecting mismatches in device attributes to prevent unauthorized access. Furthermore, service providers use this technology to identify and mitigate malicious activities, such as botnet attacks, thereby strengthening system integrity.

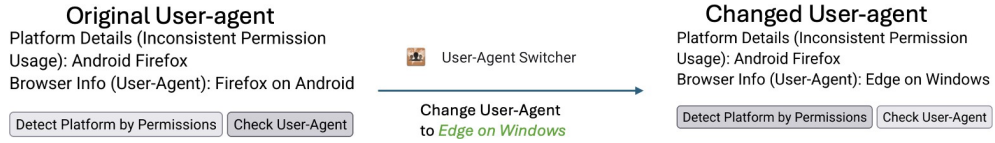

Figure 6: Screenshot of Device Identification Attack

This attack specifically targets browsers that support PWA permissions and have a significant market share (over 1% of the global browser market). The supported browsers are listed in Table 9. We assume that attackers or legitimate websites can use this technique to track user device information, such as platform details (e.g., iOS, Android, or Desktop). On Android devices, they can infer browser information. Users do not need to install the PWA; simply opening a PWA via a link allows adversaries to infer device information and use it for their purposes.

The root cause of this issue lies in the design of PWAs, where each browser independently determines how to support permissions. Unlike native apps, where permissions are standardized, PWAs rely on browser-specific implementations, leading to fragmented behaviors across platforms. This inconsistency not only degrades the user experience but also creates opportunities for attackers to infer sensitive information. By analyzing permission behaviors, attackers can deduce the user's browser and device configuration, circumventing any attempts to obfuscate such information using user-agent modifications.

As shown in Table 5, Android browsers exhibit differing default and unsupported values for certain permissions, while iOS browsers behave differently from their Android counterparts. Additionally, permission behavior varies between desktop and mobile devices. For example, the *local-fonts* permission is prompted on desktop browsers but unsupported on mobile devices. These inconsistencies provide a reliable side channel for distinguishing between devices and platforms.

Device identification shares similarities with related techniques, such as browser identification and browser fingerprinting, yet it employs distinct mechanisms. For instance, browser identification has been extensively studied in prior research. Martin et al. [57] proposed an approach to identify browsers by leveraging features of the JavaScript engine. They also demonstrated that browser information obtained from the user-agent string can be manipulated, highlighting potential vulnerabilities in this approach. Subsequent studies utilized features of HTML5 and CSS3 to detect browser-specific attributes effectively [4, 9]. Another closely related technique is *browser fingerprinting*, as outlined by Zhang et al. [72]. Browser fingerprinting enables the tracking of unique user entities by leveraging cross-browser-related fingerprint features. In essence, all three techniques exploit inconsistencies in browsers. Browser fingerprinting typically leverages features of HTML5, CSS, JavaScript, or hardware characteristics to distinguish users. While browser fingerprinting primarily focuses on identifying users, browser identification emphasizes distinguishing between browser versions. Therefore, utilizing our technique alongside existing methods provides a comprehensive and accurate picture of the user's platform, browser, and browser version.

Table 8: Comparison of Device Identification and Browser Identification on Android Browsers

| Browser | Device Identification | Browser Identification [9] |
|---|---|---|
| Chrome | Chrome on Android | Chrome on Linux |
| Samsung Internet (SI) | SI on Android | Chrome on Linux |
| Firefox | Firefox on Android | Firefox on Linux |
| Opera | Opera on Android | Chrome on Linux |
| Edge | Edge on Android | Edge on Linux |
| Brave | Brave on Android | Chrome on Linux |

Notes: The red-highlighted portions indicate incorrect platform or browser identification.

Our approach offers two key advantages over existing browser identification techniques. First, it supports execution within Service Workers, enabling device identification to function even when the PWA is closed. This capability allows for continuous background processing and facilitates the detection of users employing tools such as User-Agent changers by analyzing their IP addresses. Second, it achieves higher accuracy in distinguishing Android browsers compared to existing tools, as shown in Table 8. Existing tools often fail to differentiate browsers beyond Chrome and frequently misidentify the platform as generic Linux. Our method augments existing browser fingerprinting techniques with this enhanced feature set. Figure 6 highlights the advantages of permission-based device fingerprinting over user-agent-based methods. Unlike user-agent spoofing, permission-based fingerprinting leverages intrinsic device characteristics, such as unsupported permissions, which remain unaffected by user actions or permission state changes, ensuring reliable device identification.

## 5.3 PWA API Abuse Attack

In our study, we identified several abuse vectors stemming from the misuse of exposed PWA-related APIs, highlighting systemic weaknesses in permission enforcement across platforms. The most impactful case is the Browser Infinite Crash Attack, which exploits improper permission handling and can cause the browser to enter an unrecoverable crash loop. Once triggered, the browser becomes entirely unusable—persisting across restarts—and can only be restored by complete uninstallation and reinstallation. This attack is particularly concerning because it leverages a legitimate API, navigator.permissions.query(), and does not rely on any exploit beyond inconsistent platform-specific implementation. The vulnerability arises when a permission query references an unregistered permission type on Android, leading to a fatal internal error during permission resolution.

The threat model assumes a malicious actor who publishes a PWA embedded with crafted service worker logic. Once a user accesses the PWA, the service worker remains active in the background, enabling the attack to be triggered even without ongoing user interaction or visibility of malicious behavior in the DOM. This makes detection and prevention significantly more challenging. The importance of this attack lies in its demonstration of how subtle gaps in permission registration and platform handling—especially for newer or less-documented permissions—can lead to severe consequences. It highlights the need for more robust and consistent permission validation mechanisms across browser platforms. Technical details, root cause analysis, and code demonstrating the crash behavior are provided in Appendix A.6.1.

## 6 Root Causes of Security Risks

In this section, we analyze the root causes of the security risks introduced by the current PWA permission policy. The most fundamental issue lies in the design decision to scope PWA permissions per-site rather than per-app. This approach allows multiple PWAs under the same origin to share permission states, even if they present themselves to users as distinct applications. We investigate whether this behavior results from technical constraints in browser implementations, deliberate design trade-offs, or other considerations.

Our analysis focuses on three major browsers: Chrome, Firefox, and Safari. Other browsers such as Edge and Brave are Chromium-based and therefore exhibit permission behavior identical to Chrome. Safari on iOS implements what we refer to as an *ephemeral policy*, in which no persistent permission state is maintained. Each time a PWA is launched or a browser session is resumed, the browser re-requests permissions (with the exception of geolocation). This policy provides strong isolation between PWAs but at the cost of reduced usability due to repeated permission prompts. Firefox on Android follows an *adaptive policy*. It gives users the option to decide whether permission decisions should be remembered. If the user consents, the permission is stored persistently; otherwise, the browser mimics Safari's behavior by prompting on each use. This policy attempts to balance usability and security. In contrast, Chrome adopts a *persistent policy*, where permission decisions are stored indefinitely unless explicitly revoked by the user. While this improves usability, it introduces a critical security concern: different PWAs sharing the same origin also share the same permission state.

These inconsistencies arise because current permission models are origin-centric rather than app-centric. We argue that permissions should be scoped per-app for PWAs and per-site for traditional websites. This distinction is crucial, as PWAs are often distributed via app stores where users identify apps by name and icon—not origin. Since URLs are typically hidden in app stores, users cannot tell if two distinct-looking PWAs share the same origin and permission state. To assess the feasibility of a per-app permission model, we propose the following pseudocode for distinguishing PWAs from traditional sites and assigning permissions accordingly, as shown in Listing 1.

```
if is_https(origin) and has_web_app_manifest(
    origin):
    manifest = get_manifest(origin)
    if 'id' in manifest:
        app_id = manifest['id']
    else:
        app_id = manifest['start_url']
    assign_permission_scope(app_id)
else:
    assign_permission_scope(origin)
```

**Listing 1: Per-App Permission Scoping Algorithm**

This logic first verifies whether the origin supports PWA features (i.e., served over HTTPS and includes a manifest). If so, it extracts a unique application identifier from the manifest. The preferred field is id, with start_url as a fallback. Permissions are then scoped based on this identifier. Otherwise, permissions default to origin-based scoping. To assess the implementation implications, we examined Chromium's existing permission architecture. The following snippet from the Background Fetch permission context demonstrates how permission status is currently determined [60], as shown in Listing 2.

As shown in Listing 2, permission decisions are scoped using origin-based parameters. Our proposal suggests substituting these with an application identifier derived from the manifest, thereby enabling per-app scoping. This change is conceptually straightforward and requires minimal engineering effort. The primary challenge lies not in technical complexity but in ensuring robustness in design. For instance, if a site disables its manifest or dynamically alters its metadata, the browser may fail to recognize it as a PWA. Such scenarios can result in inconsistent permission behavior or unnecessary re-prompts, ultimately degrading user experience. These observations illustrate how current browser implementations reflect different trade-offs along the usability–security spectrum. Safari favors strong isolation at the cost of convenience, Chrome prioritizes usability with weaker isolation, and Firefox adopts a hybrid approach. These discrepancies highlight the need for a unified, app-aware permission model that provides both robust security and consistency with user expectations.

```
auto permission_result = permission_context ->
    GetPermissionStatus(
    render_frame_host,
    app_id /* was requesting_origin */,
```

```
4        app_id /* was embedding_origin */
5   );
```

**Listing 2: Origin-Based Permission Scoping in Chromium**

## 7 Mitigation Strategies

The root cause of the identified attacks lies in inconsistent and improperly designed PWA permission policies across browsers. These issues stem from the hybrid nature of PWAs, which combine the flexibility of web technologies with the behavior of standalone apps. To address these challenges, we propose targeted mitigation strategies for each attack, along with broader improvements to the PWA permission architecture.

Permission leakage occurs when multiple PWAs under the same origin share permission states, allowing one PWA to silently inherit sensitive permissions previously granted to another. This is especially problematic for powerful capabilities like NFC, where a malicious PWA can read or write to NFC tags without explicit user consent. To mitigate this, permission management should be isolated at the per-PWA level, even within the same origin. Each installed PWA must request permissions independently, preventing silent inheritance. This model resembles Android's per-app permission boundaries and would ensure stronger separation between PWAs. Additionally, permission states should not be propagated across browsers, particularly when one browser has weaker enforcement.

Device identification attacks exploit inconsistencies in permission defaults and behaviors across browsers and platforms. Subtle differences in how permissions are granted or queried can be used as side channels to fingerprint a user's device and browser configuration. To address this, we recommend unifying permission behavior across browsers wherever possible. When full standardization is not feasible, limiting programmatic access to permission states—such as restricting or obfuscating responses from *navigator.permissions.query*—can reduce the risk of fingerprinting. Injecting controlled noise or requiring user interaction before exposing permission states are also viable mitigation paths.

Abuses of the Permission API arise from its underspecified design and inconsistent enforcement across implementations. In extreme cases, attackers may exploit this to crash browsers or trigger unexpected behavior without user installation. Our responsible disclosure of such an issue led to a prompt fix by Chrome developers, highlighting the importance of vendor responsiveness. To prevent similar problems, permission interfaces should be hardened through stricter validation, and edge cases—such as background execution or undefined platform states—must be carefully audited during browser development.

Finally, to enhance transparency and restore user control, we propose a dynamic reauthorization model inspired by iOS. Under this approach, access to sensitive permissions such as geolocation, microphone, and camera must be explicitly reapproved at the beginning of each PWA session or after reinstallation. While some browsers have introduced session-based permission prompts, they are often undermined by user habituation. Dynamic reauthorization provides a stronger alternative by aligning permission access with real-time user intent and minimizing long-term overreach.

## 8 Discussion

**Limitations.** While effective in analyzing PWA permissions, our tool, *Permissioner*, has certain limitations. It relies on manual interaction due to the complexity and variability of browser behaviors, which introduces potential human error. These challenges stem from implementation differences across browsers and platforms, requiring tailored experiments for comprehensive analysis. Nevertheless, the experiments conducted with *Permissioner* are replicable and provide a solid foundation for understanding the PWA permission system. As the first comprehensive analysis of PWA permissions, this work identifies critical issues and offers insights into their implications. However, the PWA permission system is still evolving, with many APIs and descriptors remaining experimental. This presents both challenges and opportunities. While we highlight potential security risks—particularly in permissions like NFC—further research is needed to uncover additional vulnerabilities and assess the necessity of various permissions in specific PWA contexts.

## 9 Related Work

**Browser Permissions.** Nomoto et al. [58] studied permission behaviors in mobile and desktop browsers, revealing inconsistencies in prompts and enforcement. However, their work does not focus on PWAs, which introduce unique mechanisms such as service workers and installability. Kim et al. [52] analyzed browser extensions with manifest-declared permissions, improving transparency. In contrast, PWAs lack similar declarations, making their permission scopes harder to assess and audit.

**Android Permissions.** Android's permission model is often viewed as a reference for secure mobile permission design. Felt et al. [46] demonstrated that declaring permissions upfront enhances transparency and enforcement. They also introduced *permission re-delegation*, where one app leverages another's permissions to bypass user consent [47]. Almomani et al. [39] examined *collision attacks*, where loosely scoped permissions allow privilege escalation. While Android offers useful principles, applying them directly to PWAs presents challenges due to differences in platform structure and permission scope.

**Browser Fingerprinting.** Acar et al. [38] introduced FPDetective, exposing large-scale fingerprinting beyond cookie-based tracking. Englehardt and Narayanan [45] conducted a web-scale study across one million sites, confirming fingerprinting as a growing privacy threat. Laperdrix et al. [53] showed how modern APIs like Canvas and Audio enhance fingerprinting accuracy, particularly on mobile. Starov and Nikiforakis [65] quantified the fingerprintability of browser extensions, while Cao et al. [42] demonstrated that OS- and hardware-level features can enable cross-browser fingerprinting. These studies focus on traditional APIs and extensions; in contrast, our work reveals new fingerprinting vectors introduced by PWAs through permission inconsistencies and cross-platform variations—an area largely overlooked in existing research.

## 10 Conclusion

This paper systematically analyzes PWA permission management and identifies three critical attack types: permission leakage, device identification, and Permission API abuse. Using the *Permissioner* tool, we discovered implicit permission access patterns and cross-browser inconsistencies. We reported these findings to browser vendors, leading to the resolution of several vulnerabilities and substantial improvements in PWA permission security.

## A Appendix

### A.1 PWA popularity and use cases

Progressive Web Apps (PWAs) have gained substantial traction in recent years as a modern approach to delivering web-based applications with native-like capabilities. According to PWA Stats [62], many major companies have adopted PWAs due to their performance benefits, offline capabilities, and reduced development costs. For instance, Twitter Lite saw a 75% increase in tweets and a 20% decrease in bounce rate after launching its PWA [68]. Similarly, Starbucks reported that their PWA is 99.84% smaller than their iOS app and doubled the number of daily active users [64]. PWA adoption is not limited to social media or e-commerce platforms. Media organizations like Forbes [48], news outlets like The Washington Post [59], and travel platforms such as Trivago [67] have also embraced PWAs to enhance user engagement and deliver a seamless experience across devices. The surge in PWA adoption is also reflected in browser support and developer enthusiasm. Major browsers including Chrome, Edge, and Safari have rolled out increasing levels of PWA support [41], and frameworks like Angular, React, and Vue provide official PWA toolkits, enabling easier integration. These developments position PWAs as a viable alternative to native apps for many modern web services.

### A.2 PWA and TWA differences

On Android, Chrome treats a PWA as an installable application that can run in a standalone window without browser UI, offering a near-native experience. Users can add a PWA to the home screen via Chrome, and once launched, it appears as a separate task with a splash screen and its own app icon. However, this is distinct from Trusted Web Activities [44, 50], which allow developers to wrap a PWA within a native Android application and publish it on the Google Play Store.

PWAs differ significantly from traditional web applications in both user experience and system integration. Traditional web apps are accessed via the browser and run within its user interface, often with visible address bars, browser toolbars, and tab management. In contrast, PWAs are installable applications that run in standalone windows, with splash screens, app icons, and full-screen modes that resemble native apps. PWAs are also enhanced by Service Workers, enabling offline support, background sync, and push notifications — capabilities that are unavailable to traditional web apps. Additionally, PWAs benefit from manifest files that define metadata such as icons, names, and theme colors, allowing deeper integration with the host operating system.

However, despite these advantages, PWAs were historically limited in accessing system-level features compared to native apps. To address this, Google launched *Project Fugu*, an initiative aimed at closing the capability gap between PWAs and native applications [49]. Project Fugu introduces a range of powerful web APIs that extend the functionality of PWAs. For example:

- **File System Access API**: Allows PWAs to read and write files directly on the user's local file system, enabling use cases such as text editors, media processors, and IDEs.
- **Clipboard API**: Enables reading from and writing to the system clipboard, supporting complex data types beyond plain text (e.g., images and rich content).
- **Contact Picker API**: Grants PWAs permission to access the user's contact list with user consent, useful for messaging or productivity apps.
- **Web Share and Web Share Target API**: Facilitates sharing text, URLs, or files from the web app to other apps (or receiving shared content), promoting deeper system integration.
- **Notification Triggers API (experimental)**: Allows scheduling notifications at specific times, even when the app is not active — similar to native alarm-based behavior.
- **Idle Detection API**: Notifies the application when the user is idle, helping apps decide when to sync data or save state.
- **Window Controls Overlay**: Gives PWAs control over the window title bar area, enabling custom UI similar to desktop apps.
- **Screen Wake Lock API**: Prevents the device screen from dimming or locking, which is essential for media, fitness, or kiosk-style apps.

While traditional web apps remain limited to sandboxed browser environments, PWAs empowered by Project Fugu represent a new generation of web applications that closely approximate the capabilities of native apps.

### A.3 Dataset and Experiment Setup

To validate our hypothesis, we conducted experiments on major platforms supporting PWA installations, analyzing the source code of browsers with a market share greater than 1%. Table 9 summarizes PWA installation support across different platforms and operating systems, highlighting variations in compatibility. On mobile platforms, we grouped browsers into three categories for systematic investigation: (1) Chromium-based browsers (e.g., Chrome, Samsung Internet, Edge, Brave, Opera), (2) iOS-based browsers (e.g., Safari, Chrome, Firefox, Edge), and (3) Firefox on Android. Our experiments showed that browsers within the same category exhibited similar PWA permission behaviors, enabling us to identify patterns and inconsistencies effectively. On desktop, we focused on Chrome and Edge, as these were the primary browsers supporting PWA installation.

Our experimental setup utilized various devices to ensure consistency and comprehensive coverage. For Android-based experiments, we used a BLU G53 64GB Android phone and an Android Studio Pixel 8 virtual emulator. Desktop experiments were conducted on an Apple MacBook Pro M3 for macOS, while Windows and Linux experiments were performed using virtual machines. For experiments requiring Microsoft Store on Windows, we utilized an HP laptop with an Intel Core i7 processor.

The dataset for our experiments was derived from dataset [31], which provided raw data obtained from Common Crawl. While the original dataset defined PWAs more loosely and missed many relevant entries, we refined it by strictly adhering to a more precise definition of PWAs. Specifically, we required each PWA to include a manifest file with essential fields such as name and display, which are necessary for installation. In contrast, dataset [31] only checked for the presence of a manifest file, which was sometimes unrelated to PWAs. Additionally, we verified whether the PWA included a service worker, although it is not a mandatory component of PWAs, to further enhance the robustness of our experiments.

Our refined dataset is the most comprehensive and accurate dataset for PWAs in existing related work, with more stringent criteria for PWA inclusion and a broader scope of browser coverage. This approach not only improved the quality of our analysis but also ensured that our findings could reliably address the inconsistencies and risks in PWA permissions.

## A.4 Permission Descriptor Analysis

Table 15 provides a comprehensive analysis of permission descriptors, detailing their invocability, prompted status, capabilities, usage counts, and access in Service Workers (SW). From the table, we observe that most permissions are invocable, meaning they can be programmatically executed on at least one platform. However, certain descriptors, such as magnetometer, are not invocable on any platform, primarily due to unresolved security concerns or technical limitations. In contrast, permissions that are invocable on at least one platform are classified as such, even if some platforms do not yet support their invocation. For permissions that are currently non-invocable, browsers often cite security risks or other considerations, leaving developers unable to utilize them until these issues are addressed.

Regarding prompted permissions, our analysis identified 12 descriptors that trigger user prompts. These permissions typically govern access to highly sensitive capabilities, such as location, camera, and microphone, which pose direct privacy risks if misused. The explicit consent mechanism serves as a critical safeguard to ensure user awareness and control over such permissions. We also studied Service Worker access to permissions, as this represents a significant security concern. Service Workers operate in the background, allowing them to perform tasks without direct user interaction. If Service Workers were allowed to invoke sensitive permissions without user awareness, it could result in severe security vulnerabilities. Fortunately, we found no evidence of Service Workers being able to directly invoke permissions. However, Service Workers can query the status of permissions (e.g., allow or denied). This capability enables Service Workers to infer user preferences and device states, which can contribute to fingerprinting and potential privacy violations.

Several permissions warrant deeper analysis due to their importance. For instance, clipboard-write is the most frequently used permission, appearing in 32,135 PWAs. While this permission does not trigger a prompt, it allows developers to write data to the clipboard without user consent, which could be exploited for malicious purposes. On the other hand, permissions like geolocation, microphone, and camera are prompted and widely used, indicating their critical role in enhancing PWA functionality while also posing privacy risks.

Overall, our findings emphasize the need for improved differentiation and management of permissions in the context of PWAs and TWAs. The ability of Service Workers to query permission statuses highlights an area that requires further scrutiny to mitigate risks associated with user tracking and fingerprinting.

## A.5 Permission Default Value Analysis

As demonstrated in Table 10, we present the default values for all browsers and platforms included in our experiment. The symbol - indicates that the feature is not supported on the respective platform or browser, *g* signifies that the feature is granted by default, *p* denotes that the feature prompts the user by default, and *d* represents that the feature is denied by default. This table was utilized to construct the attacks discussed in Section 5.2, providing a foundational understanding of the default permission behavior across diverse environments.

The table highlights several noteworthy differences in the behavior of permission descriptors across various platforms, including iOS, Android, and desktop environments. Firstly, there are clear variations in the default states of permissions across platforms. For instance, while descriptors such as *camera*, *microphone*, and *geolocation* are prompted (*p*) by default across all platforms, descriptors like *clipboard-read* and *clipboard-write* exhibit mixed behavior. On Android, *clipboard-read* defaults to prompted in some cases but is denied (*d*) or granted (*g*) by default in others, depending on the browser implementation. Such inconsistencies can lead to divergent user experiences and potential confusion regarding when and why permissions are requested.

Secondly, certain descriptors are entirely unsupported on specific platforms. For example, *ambient-light-sensor*, *accessibility-events*, and *local-fonts* are universally unavailable (-) across all platforms and browsers. This absence is likely due to technical limitations or security concerns associated with these features. Similarly, descriptors like *nfc* are unavailable on iOS and some desktop browsers but are prompted or denied by default on Android, reflecting differences in platform capabilities and use cases.

Another striking observation is the behavior of descriptors like *background-fetch* and *background-sync*, which are granted or denied

**Table 9: PWA Installation Support Across Different OS and Platforms**

| Platform | OS | Safari | Firefox | Chrome | Edge | Opera | Brave | Samsung Internet |
|---|---|---|---|---|---|---|---|---|
| Desktop | Linux, macOS, Windows | O | O | ● | ● | O | O | † |
| Mobile | iOS | ● | ● | ● | ● | O | O | † |
|  | Android | † | ● | ● | ● | ● | ● | ● |

● : PWA installation is supported, O: PWA installation is not supported, †: Browser is not supported in this OS or platform by default on Android depending on the browser, while remaining unsupported on iOS. This disparity underscores the need for platform-specific considerations when developing PWAs to ensure consistent functionality across devices.

|  | IOS | Android | | | | | Desktop | |
|---|---|---|---|---|---|---|---|---|
| PWA Permission Descriptor | SCFE | C | SI | F | E | O | B | C | E |
| accelerometer | - | g | g | - | g | g | d | g | g |
| accessibility-events | - | - | - | - | - | - | - | - | - |
| ambient-light-sensor | - | - | - | - | - | - | - | - | - |
| background-fetch | - | g | g | - | g | d | g | g | g |
| background-sync | - | g | d | - | g | g | d | g | g |
| camera | p | p | p | p | p | p | p | p | p |
| clipboard-read | - | p | p | - | p | d | p | p | p |
| clipboard-write | - | g | g | - | g | g | g | g | g |
| display-capture | - | p | d | - | p | p | p | p | p |
| geolocation | p | p | p | p | p | p | p | p | p |
| gyroscope | - | g | g | - | g | g | d | g | g |
| idle-detection | - | p | p | - | p | d | p | p | p |
| local-fonts | - | - | - | - | - | - | - | - | - |
| magnetometer | - | g | g | - | g | g | d | g | g |
| microphone | p | p | p | p | p | p | p | p | p |
| midi | - | p | p | p | p | p | p | g | p |
| nfc | - | p | p | - | p | d | - | p | p |
| notifications | p | p | p | p | p | d | p | p | p |
| payment-handler | - | g | g | - | g | d | g | g | g |
| periodic-background-sync | - | g* | d | - | d | p | d | g | g |
| persistent-storage | - | p | p | p | p | d | p | p | p |
| push | p | p | p | p | p | p | p | p | p |
| screen-wake-lock | p | g | g | g | g | d | g | g | g |
| storage-access | - | g | p | p | g | - | - | g | g |
| system-wake-lock | - | - | - | - | - | - | - | - | - |
| top-level-storage-access | - | - | - | - | - | - | - | - | - |
| window-management | - | d | d | - | d | d | d | d | d |
| captured-surface-control | - | - | - | - | - | - | - | - | - |
| speaker-selection | - | - | - | - | - | - | - | - | - |
| keyboard-lock | - | p | - | - | p | d | p | p | p |
| pointer-lock | - | p | - | - | p | d | p | p | p |
| fullscreen | - | - | - | - | - | - | - | - | - |
| web-app-installation | - | - | - | - | - | - | - | - | - |

**Table 10: Allow by default. Notes: p means that permission is prompted state by default, g means that permission is granted by default, d means that permission is denied by default. g* means PWA allowed by default by TWA on browsers denied by default.**

**Table 11: PWA Permission Descriptors and Web APIs (Chrome-based Browsers)**

| Permission Descriptor | Web API |
| --- | --- |
| accelerometer | Sensor API |
| accessibility-events | Accessibility Object Model |
| ambient-light-sensor | AmbientLightSensor |
| background-fetch | Background Fetch API |
| background-sync | Background Synchronization API |
| camera | Media Capture and Streams API |
| clipboard-read | Clipboard API |
| clipboard-write | Clipboard API |
| display-capture | Screen Capture API |
| geolocation | Geolocation API |
| gyroscope | Sensor API |
| idle-detection | Idle Detection API |
| local-fonts | Local Font Access API |
| magnetometer | Sensor API |
| microphone | Media Capture and Streams API |
| midi | Web MIDI API |
| nfc | Web NFC API |
| notifications | Notifications API |
| payment-handler | Payment Handler API |
| periodic-background-sync | Web Periodic Background Synchronization API |
| persistent-storage | Storage API |
| push | Push API |
| screen-wake-lock | Screen Wake Lock API |
| storage-access | Storage Access API |
| system-wake-lock | Screen Wake Lock API |
| top-level-storage-access | Storage Access API |
| window-management | Window Placement API |
| window-placement | Window Placement API |
| captured-surface-control | Captured Surface Control APIs |
| keyboard-lock | Keyboard Lock API |
| pointer-lock | Pointer Lock API |
| fullscreen | Fullscreen API |
| speaker-selection | Audio Output Devices API |

Our tests revealed that the screen-wake-lock permission is granted by default in most browsers, allowing PWAs to prevent devices from entering sleep mode. Similar issues in Android native apps, such as unnecessary wakeups and wake lock leakage, have been linked to excessive battery drain and poor user experience [55]. This default behavior in PWAs indicates the potential for similar misuse, leading to resource exhaustion and security vulnerabilities. Permissions like storage-access are particularly concerning, as they enable third-party scripts to manipulate storage without user awareness. While these permissions may not appear immediately risky, they facilitate tracking across sessions and sites, raising significant privacy concerns. In contrast, Android apps mitigate such risks through stricter permission isolation, requiring explicit user consent for comparable operations.

These variations illustrate a fragmented landscape in PWA permission management, influenced by differences in platform policies, browser implementations, and underlying technical constraints. Understanding these differences is crucial for developers aiming to create secure and user-friendly PWAs. Furthermore, this analysis emphasizes the importance of unified standards to mitigate inconsistencies and improve the overall permission experience across platforms.

### A.6 Permission API Identification

In our experiment, we identified a variety of APIs that correspond to the permissions required by Progressive Web Apps (PWAs). This identification process was conducted by analyzing the source code from the Chromium project [19]. The Table 11 summarizes the permission descriptors and the corresponding web APIs we discovered. These findings are crucial for our subsequent experiments, as they provide a foundational understanding of how different permissions are implemented across Chrome-based browsers. By mapping these permissions to their respective APIs, we can better analyze their behavior and interactions. Additionally, this table serves as a valuable reference point for comparing our findings against official documentation, helping us to uncover any discrepancies or undocumented behaviors.

However, we encountered several challenges during our research. As illustrated in Figure 7, Chromium has introduced five new permission descriptors in the past year, and it is possible that more will be added in the future. Furthermore, different APIs correspond to different permissions, which are categorized in various ways. Even after identifying these APIs, practical testing is required to verify their functionality. We discovered that some APIs, despite being officially supported, are not functional in practice. These challenges highlight the dynamic and evolving nature of PWA permissions and the importance of continuous monitoring and testing.

The user actions we simulate in these experiments are categorized into four distinct types:

- *Allow:* User grants permission.
- *Deny:* User denies permission.
- *Ignore:* User ignores the permission prompt, which, depending on the browser, may automatically result in a denial after three consecutive ignores.
- *Close PWA:* We introduce this new action to capture the behavior where a user closes the PWA instead of responding to a permission prompt. This action is particularly relevant for iOS, where closing the PWA can reset previously granted permissions, thereby affecting subsequent interactions with the PWA.

Each of these user actions is essential for understanding how different browsers handle PWA permissions. Our experiments reveal that browsers typically follow a pattern where ignoring a permission prompt three times leads to an automatic denial, though the exact behavior varies across browsers. The results of these variations are detailed in the table we generated during our experiments.

The introduction of the *Close PWA* action is a critical addition, as it reflects a common user behavior not previously accounted for in permission management studies. In iOS, for example, closing the PWA can reset permissions, a unique behavior that we define and analyze as part of our study.

The primary goal of these dynamic experiments is to uncover the differences in how PWA permissions are managed across platforms,

| Matt Reichhoff | 2dbcfbd | 2023-01-03 17:48:33 | [diff] [blame] | 39 | `"top-level-storage-access",` |
| Elad Alon | dac5de9d | 2023-12-11 15:54:06 | [diff] [blame] | 40 | `"captured-surface-control",` |
| Sunggook Chue | 3d430a4 | 2024-02-16 21:27:42 | [diff] [blame] | 41 | `"speaker-selection",` |
| Takumi Fujimoto | b5aa3ef | 2024-05-10 13:10:36 | [diff] [blame] | 42 | `"keyboard-lock",` |
|  |  |  |  | 43 | `"pointer-lock",` |
| Mike Wasserman | 740e873 | 2024-06-19 19:39:19 | [diff] [blame] | 44 | `"fullscreen",` |

**Figure 7: New permission descriptors introduced in Chromium over the past year**

focusing on two key perspectives: the isolation of permissions and the mechanisms by which permissions are granted. By simulating real-world interactions, we aim to provide a comprehensive understanding of the PWA permission system, highlighting potential security risks and usability challenges that may arise in different browser environments.

*A.6.1 Permission API Abusing Attacks* As discussed in Section 4, the Permission API plays a central role in the PWA permission system. While TWA permissions, such as those for the camera and microphone, have been extensively studied, newly introduced permissions often lack clear definitions and specifications. This ambiguity leads browser vendors to overlook potential edge cases, resulting in security vulnerabilities. For instance, as demonstrated in Appendix A.7, the Sensor API exhibits no frequency limitations compared to the Android permission system. Additionally, we identified a critical issue: a Browser Infinite Crash Attack related to Chrome on Android. This attack was severe enough that Chrome addressed the vulnerability within a week of our report and requested validation of the fix.

The Browser Infinite Crash Attack exploits a vulnerability wherein a user visiting an attacker-controlled PWA experiences perpetual browser crashes. This issue persists even after the browser is closed and reopened, rendering the browser unusable. The only resolution is to completely uninstall and reinstall the browser. In this attack, the adversary embeds malicious logic in service workers rather than the main DOM, ensuring that users cannot easily identify the source of the crash. The attack is triggered when the *keyboard-lock* permission is queried. On Chrome for Android, this query causes the browser to crash immediately. Upon reopening, user interactions, such as entering a URL, are disabled, effectively locking the user out. The corresponding code is presented in Listing 3.

Through communication with browser vendors, we identified the root cause of the issue as an oversight in platform-specific permission handling. While permissions are correctly registered in desktop environments, the *keyboard-lock* permission is not registered for Android in *content_settings_registry.cc*. As a result, querying this permission triggers an unregistered *ContentSettingsType*, causing a forced *CHECK* in *host_content_settings_map.cc*, which leads to a browser crash [6]. This finding underscores a lack of consistency in permission definitions across platforms. Browser vendors acknowledged that they had overlooked the possibility of this permission being queried on Android devices. This oversight further highlights the lack of clear and comprehensive specifications in the PWA permission system. If browser vendors had recognized the unique characteristics of the PWA permission system and accounted for corner cases, this issue could have been prevented. Although the vulnerability was addressed starting with Chrome version 131, users of earlier versions, such as Chrome 128, 129, and 130, remained vulnerable to this attack.

```
1  navigator.permissions.query({ name: 'keyboard-lock' })
```
**Listing 3: Code triggering the infinite crash in Android Chrome**

## A.7 PWA Sensor API Abuse Attack

This subsection highlights the vulnerabilities in PWA Sensor API management, where implicit permissions enable high-frequency sensor data collection—reaching up to 5,000 Hz—without requiring any user consent. In contrast to Android, which imposes frequency caps and requires explicit permission declarations, PWAs lack such safeguards. As a result, attackers can leverage unrestricted sensor access to collect high-resolution data, infer sensitive user behavior, and degrade device performance, all while bypassing traditional permission prompts.

A malicious PWA can begin accessing sensor data immediately after the user visits or installs it. Since the permission model is implicit, no additional user interaction is needed to initiate or maintain data collection. This allows attackers to continuously monitor real-time sensor outputs, such as motion or orientation, and infer private user actions. The ongoing nature of this access, combined with the absence of user notifications, makes the attack difficult to detect and particularly effective.

Sensor-based inference attacks have also been demonstrated in the Android ecosystem, where researchers showed that motion and orientation data can be used to reconstruct speech patterns, detect keystrokes, and track user movements. However, Android mitigates these risks through explicit permission requirements and frequency limitations, capping sensor access at 200 Hz. PWAs, on the other hand, are not subject to these restrictions and can operate at frequencies up to 5,000 Hz. This discrepancy significantly increases the resolution and fidelity of the collected data, thereby amplifying the privacy and security risks. Table 12 presents a comparison between PWAs and Android apps in terms of sensor access control and platform-level safeguards.

The implications of unrestricted, high-frequency sensor access are severe. Previous research has shown that operating sensors above 200 Hz can drastically shorten battery life—from years to days under normal usage, and from days to mere hours under continuous access conditions. Furthermore, Android apps are subject to app store reviews and explicit permission flows, which act as additional layers of oversight. In contrast, PWAs bypass these protections entirely, exposing users to persistent and high-impact sensor exploitation with minimal platform scrutiny.

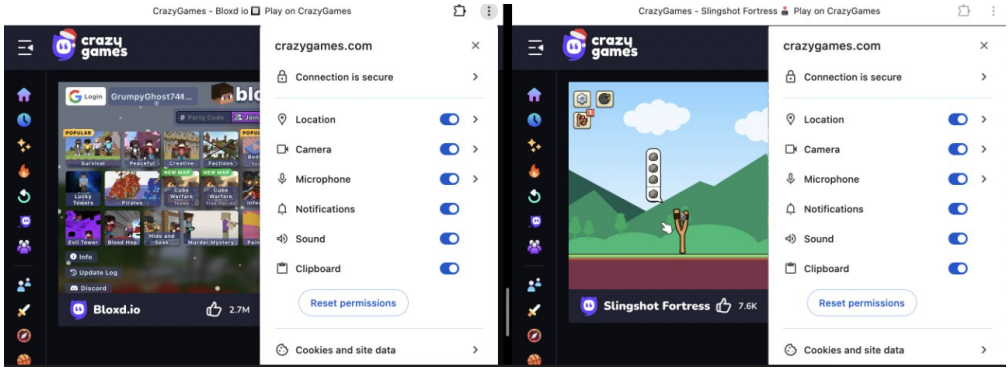

Figure 8: Example of permission sharing in *CrazyGames*

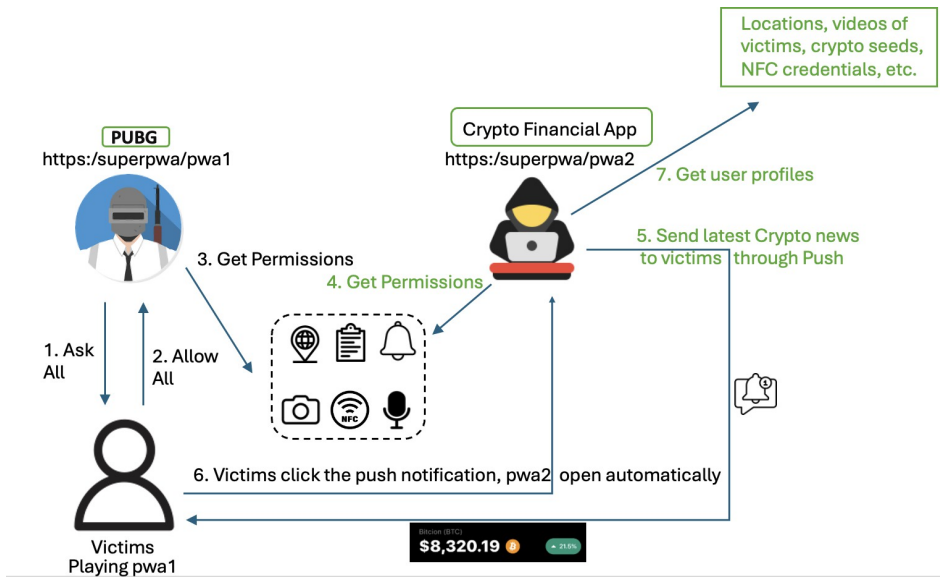

Figure 9: Illustration of the PWA Permission Leakage Attack

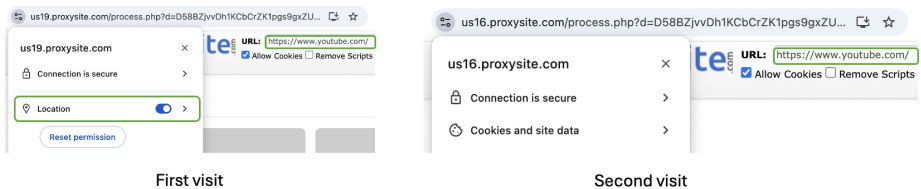

Figure 10: Example of rehosting website permission isolation

Although adoption of the Sensor API among current PWAs remains limited, its unregulated use poses critical risks. Some legitimate PWAs, such as *gabriel-saraiva.com* (60 Hz) and *madisonsolutions.co.uk* (40 Hz), demonstrate responsible usage of the gyroscope to improve user experience. However, inconsistencies across platforms and devices reduce user awareness and hinder detection, allowing attackers to selectively exploit sensors on compatible hardware. The lack of platform-enforced restrictions means that attackers can not only degrade device performance and deplete batteries, but potentially cause hardware stress or damage through prolonged high-frequency usage. It is important to note that while browsers do not impose limits on Sensor API frequency, hardware capabilities may inherently constrain maximum access rates. In our experiments conducted on a budget Android smartphone, the gyroscope was limited to 60 Hz due to hardware design.

**Table 12: Comparison of Sensor API Access and Impact: PWAs vs. Android Apps**

| Aspect | PWAs | Android Apps |
|---|---|---|
| Permission Requirement | No explicit user consent required | Explicit consent via app manifest |
| Sensor Frequency Limit | Up to 5,000 Hz | Capped at 200 Hz |
| Battery Life Impact | From years to a few days (5,000 Hz) | Minimal due to frequency limitation |
| Oversight and Safeguards | No review process; unrestricted access | Reviewed during app store submission |

## A.8 PWA Permission Leakage attacks results

In this section, we demonstrate the additional capabilities enabled after launching these attacks and what attackers could potentially acquire in real-world scenarios through controlled environment experiments. The experiments were conducted within a local network, ensuring no harm was caused to any external individuals or systems.

*A.8.1 Examples of Clipboard and Recording Capabilities* Figure 2 shows a screenshot of a PWA permission leakage attack, where sensitive information, such as a user's crypto seed and IP address, is disclosed without user consent. This allows attackers to potentially steal cryptocurrency by exploiting leaked seed information. Figure 3 illustrates unauthorized video recording, where user footage is secretly captured and sent to a remote server without permission.

*A.8.2 Examples of NFC Capabilities* NFC tags can be read or written to without user interaction when the device is in close proximity. As shown in Table 13, real-world use cases of NFC include Wi-Fi configuration sharing, smart home control, and transportation ticketing. In our dataset, we identified three PWAs using the NFC API, but their specific use cases were unclear due to login requirements. However, we confirmed that https://www.leedo.org uses NFC for business card management, further demonstrating the practical value of NFC tags, as shown in Figure 4.

*A.8.3 TWA and PWA Engagement Rate Discussion* Click-Through Rate (CTR) is a key metric used to measure user engagement. It represents the percentage of users who click on a link or take an action compared to the total number of users who view it. A higher CTR often reflects greater user interest and the effectiveness of the content or strategy in capturing attention. For TWAs, engagement rates remain relatively low, with CTRs generally not exceeding 10% [1]. In contrast, PWAs demonstrate higher engagement potential, particularly when using push notifications. Table 14 summarizes key factors influencing CTRs and their respective impacts.

**Table 13: Real-World NFC Use Cases in PWAs**

| Use Case |
|---|
| **Wi-Fi Sharing:** Sharing SSID and passwords through NFC tags. [2, 17, 35] |
| **Information and Anti-Counterfeit:** Embedding product details, business card information, poster details, and protecting products from counterfeiting. [3, 18, 33] |
| **Smart Home Control:** Using NFC tags to control lights and appliances. [2, 17, 35] |
| **E-Ticketing:** Storing e-tickets for buses and trains. [12, 17, 35] |
| **Luxury and Consumer Engagement:** Ensuring authenticity of luxury items and enhancing customer experience with additional product information. [10, 16] |

As shown in Table 14, strategies such as customization, actionable CTAs, and industry-specific timing can significantly enhance CTRs. When combined with rich media formats, engagement rates can potentially exceed 80% [14, 28, 29, 34].

**Table 14: Click-Through Rates (CTR) and Influencing Factors**

| Factor | CTR/Impact |
|---|---|
| Simple push notifications | 15% [28] |
| Customized notifications | +15% [28] |
| Actionable CTAs | +25% [28] |
| Use of emojis | +20% [14] |
| Industry-specific timing | +40% [14] |
| Rich media formats | +25% [14] |
| Notifications with images | +56% [34] |

**Table 15: List of Permission Descriptors, Invocability, Prompted Status, Capabilities, Counts and Access in Service Workers**

| Permission Descriptor | Invocable | Prompted | Capability | Count | SW Access |
|---|---|---|---|---|---|
| clipboard-write | ✓ | ✗ | Write text to clipboard | 32,135 | ✗ |
| clipboard-read | ✓ | ✓ | Read text from clipboard | 24,753 | ✗ |
| geolocation | ✓ | ✓ | Access location | 11,350 | ✗ |
| background-sync | ✓ | ✗ | Sync in background | 10,456 | ✓ |
| notifications | ✓ | ✓ | Send notifications | 8,691 | ✗ |
| fullscreen | ✓ | ✗ | Full-screen mode | 5,336 | ✗ |
| microphone | ✓ | ✓ | Microphone access | 2,970 | ✗ |
| camera | ✓ | ✓ | Camera access | 2,959 | ✗ |
| storage-access | ✓ | ✗ | Third-party storage access | 673 | ✗ |
| display-capture | ✓ | ✗ | Screen sharing | 539 | ✗ |
| pointer-lock | ✓ | ✗ | Pointer lock | 201 | ✗ |
| screen-wake-lock | ✓ | ✗ | Prevent screen sleep | 154 | ✓ |
| payment-handler | ✓ | ✓ | Process payments | 113 | ✗ |
| periodic-background-sync | ✓ | ✗ | Periodic data sync | 71 | ✓ |
| persistent-storage | ✓ | ✗ | Persistent storage | 53 | ✗ |
| midi | ✓ | ✓ | Access MIDI devices | 27 | ✗ |
| idle-detection | ✓ | ✓ | Detect user idle | 8 | ✗ |
| window-management | ✓ | ✓ | Manage browser windows | 8 | ✗ |
| local-fonts | ✓ | ✓ | Access local fonts | 5 | ✗ |
| nfc | ✓ | ✓ | Interact with NFC devices | 4 | ✗ |
| push | ✓ | ✓ | Push notifications | 3 | ✓ |
| gyroscope | ✓ | ✗ | Access gyroscope data | 2 | ✗ |
| background-fetch | ✓ | ✗ | Background resource fetch | 1 | ✓ |
| captured-surface-control | ✓ | ✗ | Control captured media | 1 | ✗ |
| keyboard-lock | ✓ | ✗ | Lock specific keyboard keys | 0 | ✗ |
| accelerometer | ✓ | ✗ | Access accelerometer data | 0 | ✗ |
| speaker-selection | ✓ | ✗ | Manage audio output | 0 | ✗ |
| magnetometer | ✗ | - | - | 0 | ✗ |
| accessibility-events | ✗ | - | - | 0 | ✗ |
| system-wake-lock | ✗ | - | - | 0 | ✗ |
| top-level-storage-access | ✗ | - | - | 0 | ✗ |
| ambient-light-sensor | ✗ | - | - | 0 | ✗ |
| web-app-installation | ✗ | - | - | 0 | ✗ |

**Table 16: Permission Options for Various Browsers and Permissions**

| Browser | Geolocation | Notification | Microphone | Camera | Clipboard Read |
|---|---|---|---|---|---|
| Safari (iOS) | 2 | 2 | 2 | 2 | 1 |
| Chrome (iOS) | 2 | 2 | 2 | 2 | 1 |
| Firefox (iOS) | 3 | 3 | 3 | 3 | 1 |
| Edge (iOS) | 2 | 2 | 2 | 2 | 1 |
| Chrome (Android) | 2 | 2 | 2 | 2 | 2 |
| Samsung Internet (Android) | 2 | 2 | 2 | 2 | 2 |
| Firefox (Android) | 2 | 2 | 2 | 2 | 1 |
| Edge (Android) | 2 | 2 | 2 | 2 | 2 |
| Opera (Android) | 2 | 2 | 2 | 2 | 0 |
| Brave (Android) | 4 | 4 | 4 | 4 | 4 |

Note: Opera (Clipboard Read) is set to 0 because it is denied by default. Clipboard Read has 1 option on some browsers because it only has one prompt with *Paste*. 2 options represent allow or deny. Brave has 4 permission options:
- Until I close this site
- For 24 hours
- For 1 week
- Forever